\documentclass[12pt,preprint]{aastex}

\shorttitle{WD cooling sequences}
\shortauthors{Salaris et al.}

\begin{document}

\def\subr #1{_{{\rm #1}}}


\title{A large stellar evolution database for population synthesis studies: VI. 
White dwarf cooling sequences}

\author{ M.\ Salaris\altaffilmark{1},   
         S.\ Cassisi\altaffilmark{2},
         A.\ Pietrinferni\altaffilmark{2},
         P.M.\ Kowalski\altaffilmark{3}, and
         J.\ Isern\altaffilmark{4,5} 
         }

\altaffiltext{1}{Astrophysics Research Institute, Liverpool John Moores
University, 12 Quays House, Birkenhead, CH41 1LD, UK; ms@astro.livjm.ac.uk}

\altaffiltext{2}{INAF-Osservatorio Astronomico di Collurania,
via M. Maggini, 64100 Teramo, Italy;
cassisi, adriano@oa-teramo.inaf.it}

\altaffiltext{3}{Helmholtz-Centre Potsdam - GFZ German Research Centre for Geosciences,
Section 3.3, D-14473, Potsdam, Germany}

\altaffiltext{4}{Institut de Ci\'encies de l'Espai (CSIC), Facultat de Cie\'ncies, campus UAB, 08193 Bellaterra, Spain}

\altaffiltext{5}{Institut d'Estudis Espacials de Catalunya, c/Gran Capita\' 2-4, Edif. Nexus 104, 08034 Barcelona, Spain}

\begin{abstract}
We present a new set of cooling models and isochrones for both H- and He-atmosphere white dwarfs, incorporating 
accurate boundary conditions from detailed model atmosphere calculations, and carbon-oxygen chemical abundance 
profiles based on updated stellar evolution calculations from the BaSTI stellar evolution archive - 
a theoretical data center for the Virtual Observatory. We discuss and quantify the uncertainties 
in the cooling times predicted by the models, arising from the treatment of mixing during the central H- and He-burning phases, 
number of thermal pulses experienced by the progenitors, progenitor metallicity 
and the $^{12}C(\alpha,\gamma)^{16}O$ reaction rate. The largest sources of uncertainty turn out to be 
related to the treatment of convection during the last stages of the progenitor central He-burning phase, and 
the $^{12}C(\alpha,\gamma)^{16}O$ reaction rate.

We compare our new models to previous calculations performed with the same stellar evolution code, and  
discuss their application to the estimate of the age of the solar neighborhood, and the interpretation of the observed 
number ratios between H- and He-atmosphere white dwarfs.
The new white dwarf sequences and an extensive set of white dwarf isochrones that cover a large range of ages and progenitor 
metallicities are made publicly available at the official BaSTI website.  
\end{abstract}


%
\section{Introduction}
\label{intro}
%

The interpretation of photometric and spectroscopic observations of stellar populations relies on 
the use of grids of stellar models and isochrones, that have to cover a wide range of 
initial chemical compositions, stellar masses and evolutionary phases. 
The BaSTI (a Bag of Stellar Tracks and Isochrones) project\footnote{Official website at 
\url{http://www.oa-teramo.inaf.it/BASTI}} 
started in 2004 has delivered, to date, an homogeneous database of stellar evolution models, 
isochrones and integrated 
spectra for single-age, single-metallicity populations, 
encompassing a large chemical composition range appropriate for stellar populations harboured in 
star clusters and galaxies of various morphological types (Pietrinferni et al.~2004, 2006, 2009, 
Cordier et al.~2007, Percival et al.~2009). Results from BaSTI projects have been used by a large number of authors 
to address very diverse astrophysical problems like, among others, fitting eclipsing binary systems in the mass-radius plane, 
determining the ages of star clusters from their color-magnitude-diagrams (CMDs) or comparing integrated colors of 
elliptical galaxies with theoretical predictions.

BaSTI models and isochrones in their present form cover all relevant evolutionary phases  
until either the end of the thermal pulse regime along the Asymptotic Giant Branch (AGB), 
or central carbon ignition for masses without electron degenerate carbon-oxygen (CO) cores. 
In this paper we extend the evolutionary phase coverage of our database to include cooling models of 
CO-core White Dwarfs (WDs), the final evolutionary phase of stars with initial masses smaller than 
about 6-7 $M_{\odot}$. 

During the last two decades observations and theory have improved to a level that has made finally 
possible to employ WDs for determining 
ages of the stellar populations in the solar neighborhood (e.g., Winget et al.~1987, Garcia-Berro et al.~1988, Wood 1992, 
Oswalt et al.~1996), and in the nearest open (e.g., Richer et al.~1998, von Hippel~2005, Bedin et al.~2008, 2010) 
and globular (e.g. Hansen et al. 2004, 2007, Bedin et al.~2009) clusters.
Methods to determine stellar population ages from their WD cooling sequences are usually based on the comparison of 
either the observed WD luminosity function (LF - star counts as a function of magnitude, e.g.,Winget et al.~1987, 
Bedin et al.~2010) or the actual bidimensional 
WD distribution in the CMD, with their theoretical counterparts 
(see, e.g., Hansen et al.~2007). Both techniques rely on an extensive use of grids of WD cooling sequences. 

The sets of WD models largely employed in the more recent investigations 
on the age of WDs in Galactic stellar populations 
are those by Hansen~(1999, hereafter H99) and Salaris et al.~(2000, hereafter S00), computed with completely independent 
evolutionary codes and largely independent input physics. Additional recent large sets of WD evolutionary cooling models 
can be found in Althaus \& Benvenuto~(1998 -- and later updates from the same group, who has also produced extensive 
libraries of He-core WD models, presented in Serenelli et al.~2002 and Althaus et al.~2009) and Fontaine, 
Brassard \& Bergeron~(2001). 
The new grid of cooling models we present here to extend the evolutionary phase coverage of BaSTI,   
is an update of the results by S00. We include a complete set of both H- and He-atmosphere WD models (and isochrones, 
for a range of progenitor initial chemical compositions) that   
take advantage of the updated CO stratifications obtained from BaSTI AGB models, 
and employ boundary conditions from new sets of calculations of WD H and He atmospheres. 
Along with the presentation of our new models, 
we will discuss critically how WD cooling times are affected by the progenitor 
metallicity, uncertainties on the current estimate 
of the $^{12}C(\alpha,\gamma)^{16}O$ 
reaction rate and treatment of convection during the progenitor evolution. A similar analysis 
(albeit with several differences in the details) can be found in Prada-Moroni \& Straniero~(2002),   
without taking into account CO phase separation upon crystallization in the WD cooling models.

The paper is structured as follows. Section~2 presents briefly the updates in the input physics compared to S00, 
and discusses critically our choices for 
the core chemical stratifications. Section~3 analyzes the main properties of the cooling models and  
WD isochrones, shows comparisons with S00 calculations and an example of application to 
study WDs in the solar neighborhood. A summary follows in Sect.~4.

%
\section{Input physics, core and envelope stratification}
\label{inputs}

The cooling code employed in our calculations is the same described in S00, and the 
reader is referred to that paper for more details about the model input physics. 
The only differences compared to S00 calculations involve  boundary conditions, the core and envelope 
chemical compositions, and will be described below. 

We provide WD cooling models for masses equal to 0.54, 0.55, 0.61, 0.68, 0.77, 0.87, and 1.0~$M_{\odot}$, 
as in S00, plus WD isochrones, both neglecting and including the release of 
gravitational energy associated to the phase separation of the CO mixture upon crystallization 
(e.g., Stevenson~1977, Mochkovitch~1983, Garcia-Berro et al.~1988, 
Segretain et al.~1994, Montgomery et al.~1999, Isern et al. 2000 and references therein). The effect of phase 
separation is calculated as in S00. 

The range of WD masses presented here ensures a good coverage of the full spectrum of WD masses 
derived from semiempirical progenitor-WD (initial-final) mass relationships 
(see, e.g., Salaris et al.~2009).

Similar to S00, for each WD mass an initial model was converged at log($L/L_{\odot}$)$\sim$1.0--1.5 
by considering a reference CO stratification in the core, and 
a reference thickness and chemical composition of the envelope layers.

\subsection{Envelope chemical stratification}

We have computed WD models considering both pure H and pure He atmospheres. 
The H-atmosphere WD models 
have 'thick' H layers, as in H99, S00 ( andt he H-atmosphere models by Fontaine et al.~2001); 
the envelope consists of a H-layer with mass fraction $q_{H}=10^{-4} M_{WD}$ 
on top of a He-layer of mass $q_{He}=10^{-2} M_{WD}$.
The He-atmosphere WD models  
have a He-envelope with mass fraction $q_{He}=10^{-3.5} M_{WD}$, as in 
H99. With these choices for $q_{H}$ and $q_{He}$ , 
the surface convective regions that develop during the cooling process are not able 
to cross the H-He interface in H-atmosphere models, or the He-CO interface in He-atmosphere models. 

Rosseland low-temperature ($T<$10000~K) opacities 
by Alexander et al.~(1997) for a pure-He composition are employed in the envelope calculations of 
He-atmospheres models, while the Saumon \& Jacobson~(1999) low-temperature opacities are employed 
for the hydrogen envelopes , as in S00. At higher temperatures we employed the 
OPAL (Iglesias \& Rogers~1993) radiative opacities for the 
appropriate chemical composition.

As mentioned before, our envelopes have zero metal content, even though progenitor models 
have non-zero intial metallicity. Given the high efficiency of atomic diffusion 
at the very beginning of the cooling sequence (e.g., Koester~2009) all metals in the 
WD envelopes have settled above the core boundary. To have an approximate estimate of this effect 
on our H-atmosphere cooling models, we have computed the evolution of a 0.61$M_{\odot}$ model (for the reference 
CO stratification discussed in Sect.~\ref{profili}) keeping the hydrogen layers 
metal free, but considering a metal mass fraction Z=0.0198 -- the initial 
solar metal abundance in the BaSTI solar model (Pietrinferni et al.~2004) -- 
distributed uniformly throughout the underlying He-envelope.
Comparison of the cooling times with the Z=0 case in the whole envelope shows differences by typically less than 1\%.
In case of the He-atmosphere cooling models, a rough estimate of the same effect has been derived by 
calculating the evolution of a 0.61$M_{\odot}$ model with metals 
(initial metallicity Z=0.0198) redistributed uniformly in the deeper envelope layers where the pressure is 
more than 6 orders of magnitude higher than the photosphere. This means that the 
photosphere and outer envelope are still metal free, as in the reference case.
We have found differences in the cooling times by at most $\sim$2 \%, compared to the metal free case.

As a final comment, we remark that the thickness of 
the envelope layers in our models is the same for the whole range of WD masses. Also any possible dependence 
on the progenitor metallicity will be disregarded. Results from evolutionary models 
that follow the evolution throughout the AGB phase until the WD cooling sequence 
(see, e.g., Iben \& MacDonald~1986) show a dependence of $q_{H}$ and $q_{He}$
on progenitor mass and initial chemical composition, but the uncertainties in the theoretical predictions 
of the WD envelope thickness are still large, i.e.,  
they depend crucially on the precise description of the mass loss events during the TP phase.  
Calculations with constant thickness of the envelope layers provide a useful reference 
baseline grid of WD models that allow to disentangle in a more direct way the effect of 
the existing uncertainties related to core and envelope physical and chemical properties. 

For our choice of $q_{H}$ and $q_{He}$ and the starting luminosities of our calculations, 
H-burning at the bottom of the H-envelope is negligible in all but the more massive models, 
and does not affect cooling times and the value of $q_{H}$. 
For the more massive models, we have estimated with test calculations 
that the effect of H-burning on the cooling times amounts 
to a few percent. In order to keep $q_{H}$ at a strictly constant value for the whole model grid, 
we have inhibited nuclear burning in all calculations discussed in this paper.

\subsection{Boundary conditions}

The surface boundary conditions needed to integrate the stellar structure (P and T at $\tau$ = 100, 
where the diffusion approximation is valid and one can safely start to integrate the full set of stellar 
structure equations using Rosseland mean opacities) were obtained for $T_{eff}<10000$~K 
from detailed non-gray model atmospheres. These include the latest physical
improvements in the calculation of the chemistry, opacity, radiative
transfer, and equation of state of
dense hydrogen and helium in WD atmospheres (Kowalski \& Saumon
2004; 2006; Kowalski 2006a; 2006b, Kowalski et al 2007). 
At higher $T_{eff}$, 
the boundary conditions have been 
calculated by integrating a gray $T(\tau)$ relationship. As discussed in S00 (see also H99) in this 
temperature regime a gray $T(\tau)$ integration is a suitable choice.

\begin{figure}
\epsscale{1.00}
\plotone{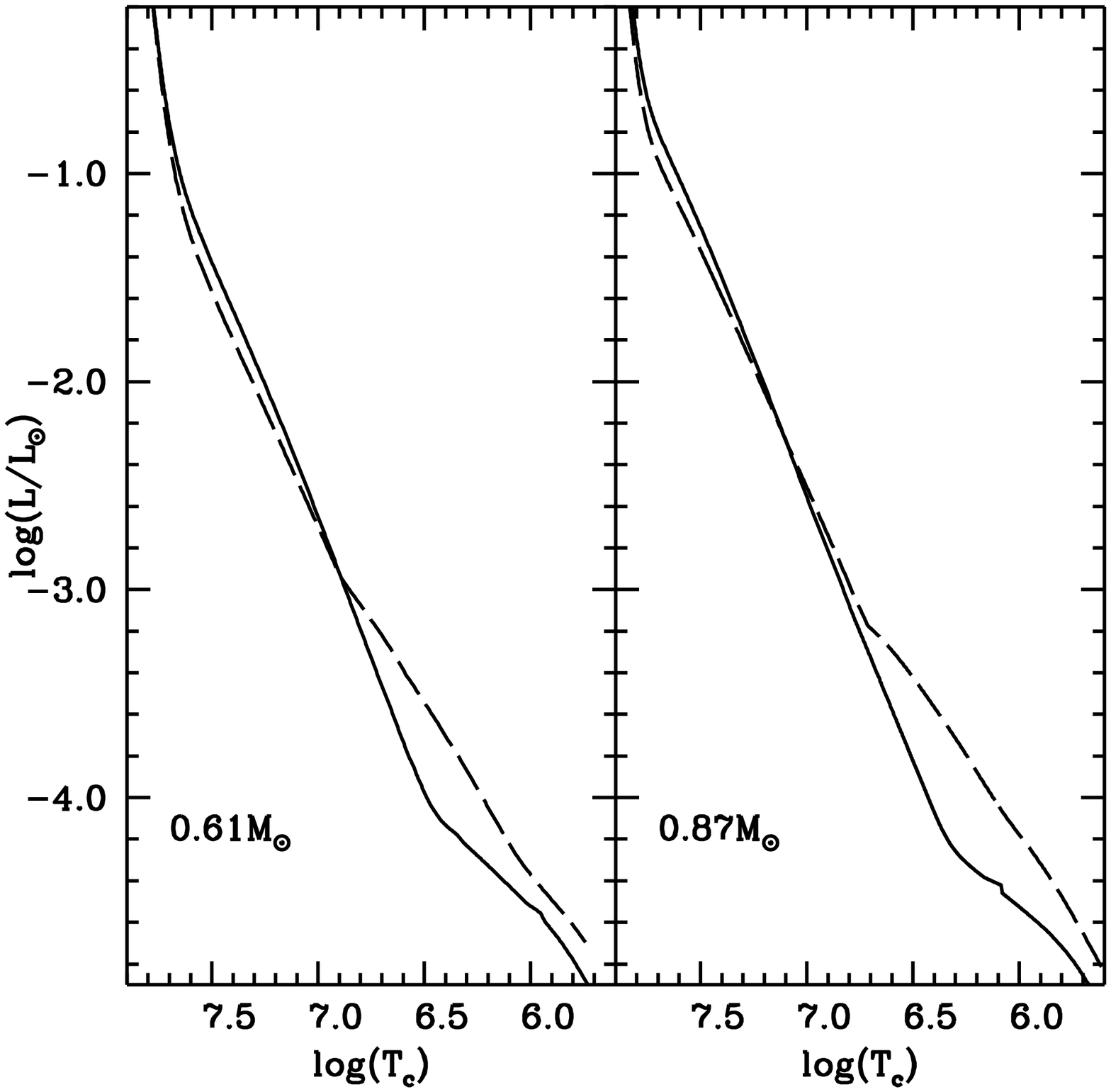}
\caption{$L-T_c$ relationships for our 0.61 and 0.87 $M_{\odot}$ WD models (without phase separation). 
Solid lines denote H-atmosphere models, dashed lines He-atmosphere ones.}
\label{LTc}
\end{figure}

Figure~\ref{LTc} compares the luminosity-central temperature ($L-T_c$) relationships for two representative 
0.61$M_{\odot}$ and 0.87$M_{\odot}$ cooling models, with both H- and He atmospheres. 
As discussed in S00 and H99, the $L-T_c$ relationship at fixed $M_{WD}$ depends 
only on the properties of the non-degenerate envelope and atmosphere, and is independent of the 
CO stratification and the treatment of crystallization 
(we display in the figure the results without phase separation). 
In case of the 0.61$M_{\odot}$ model, the He-envelope/atmosphere 
has a higher opacity (the same $T_c$ is reached at lower luminosities) 
between $\log(L/L_{\odot})\sim -1$ and $\sim -$3.0. 
Below log$(L/L_{\odot})\sim -3.0$ the two relationships  
start to diverge considerably (see also Fig.~8 in H00), the He-envelope external layers becoming 
sizably less opaque when the atmospheres become neutral (H99). At lower luminosities they tend to move back close 
to each other.
As for the 0.87$M_{\odot}$ model, the behavior is very similar. The luminosity at which the He-envelope  
becomes less opaque is shifted to slightly higher values, compared to the 0.61$M_{\odot}$ model.

As a test, we have also compared our new H-envelope 
$L-T_c$ relationship with the S00 one for the same 0.61$M_{\odot}$ mass, and did 
not find any difference between the two models.

\subsection{CO profiles}
\label{profili}
The choice of the CO stratification is extremely important, given that the rate of cooling is determined, 
among other factors, by the 
ionic specific heat, which depends on the relative proportions of carbon and oxygen. 
The additional source of energy provided by the 
crystallization process is also greatly affected by the CO profile (see, e.g., Salaris et al.~1997).

Our adopted reference CO stratifications have been obtained from the grid of BaSTI scaled solar stellar 
evolution models. More in detail, we employed the sets of models with the initial 
solar metal abundance (Z=0.0198), that include convective core 
overshooting during the Main Sequence, for it appears to be necessary to reproduce the CMD and star counts in 
young and intermediate age star clusters (see, e.g., the discussion in Pietrinferni et al.~2004 and references therein). 
Core mixing during central He-burning is treated by including semiconvection according to the method 
described by Castellani et al.~(1985). The breathing pulses occurring during the last portion of core He-burning 
have been inhibited following the method by Caputo et al.~(1989). 
The reader is referred to, e.g., Cassisi et al.~(2001), and Cassisi, Salaris, \& Irwin~(2003) for a discussion about 
breathing pulses and observational constraints on their efficiency.

\begin{figure}
\epsscale{1.00}
\plotone{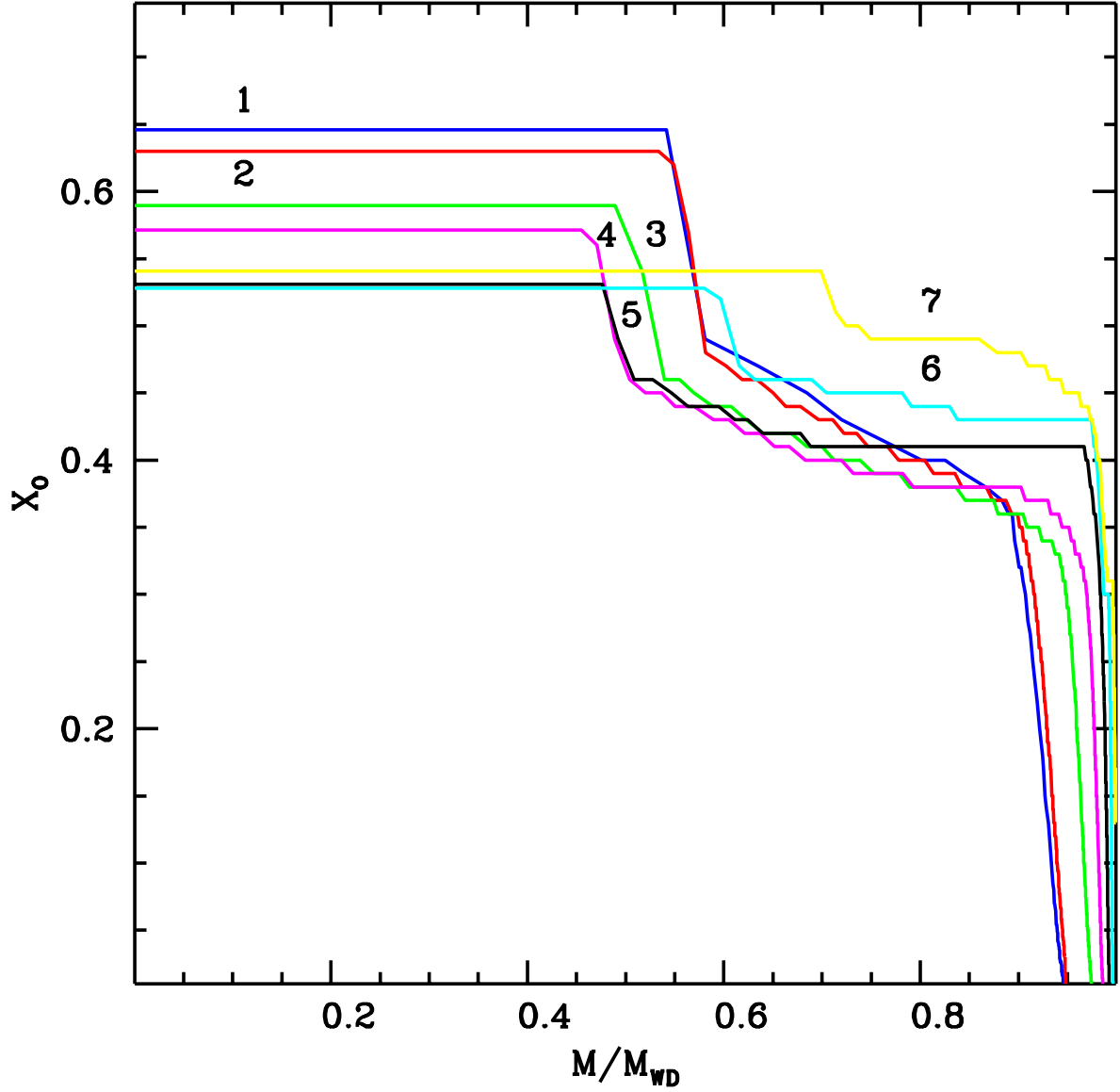}
\caption{Reference oxygen stratification (in mass fraction) for our WD models. 
Labels denote, in order of increasing number, 
the abundances for the 0.54, 0.55, 0.61, 0.68, 0.77, 0.87 and 1.0 $M_{\odot}$ models, respectively.}
\label{profilesall}
\end{figure}

For a given value of $M_{WD}$, we 
have taken the core stratification at the first thermal pulse from the progenitor model whose mass internal to the He-H 
discontinuity is equal to $M_{WD}$. 
If for some of our selected WD masses there is    
no model in the original BaSTI grid that displays the appropriate core mass at the first thermal pulse, we have performed 
additional calculations. 
All our reference chemical profiles are displayed in  Fig.~\ref{profilesall}, and are kept fixed in all 
WD isochrones we will make available in BaSTI, that span a large range of progenitor metallicities. 
This means that, when we compute WD isochrones 
for a stellar population of a given intial composition, only as far as the WD initial chemical 
stratification is concerned, 
we are neglecting the effect of the progenitor metallicity 
and we are assuming an initial-final mass relationship (IFMR) given by the core masses at the first thermal pulse of models 
including core overshooting during the main sequence.
We notice here that with our reference CO stratifications, the onset of the 
convective coupling with electron degenerate layers 
-- a process discussed in detail by Fontaine et al.~(2001) --  overlaps in time with the  
crystallization of the CO core, for the full range of WD masses.

In the following, we will evaluate the impact of these assumptions 
about the CO profile on the cooling times of a 0.61$M_{\odot}$ WD model 
(we performed the same 
analysis for a 1.0$M_{\odot}$ model, and obtained very similar results) in comparison with 
the effect of two major intrinsic uncertainties in the modelling 
of the central He-burning phase, namely the method for breathing pulse suppression, and the 
$^{12}C(\alpha,\gamma)^{16}O$ reaction rate. 
To do so, we take advantage of additional calculations for the progenitor evolutions, performed for this work. 

\subsubsection{Main sequence core overshooting}

Figure~\ref{profiles} displays several O-abundance profiles (in mass fraction) for a 0.61$M_{\odot}$ WD model, 
calculated under different assumptions, all taken after the rehomogeneization by 
Rayleigh-Taylor instability discussed in Salaris et al.~(1997). 
The two dashed lines correspond to progenitor models at the first 
thermal pulse computed with 
(our reference choice -- the profile with a very slightly higher central oxygen abundance) and without 
core overshooting on the main sequence. The two stratifications are almost identical, in spite of the fact that 
the progenitor mass 
is $\sim$3.0$M_{\odot}$ for the model with main sequence core overshooting, and $\sim$3.5 $M_{\odot}$ 
for the model without overshooting. 
The reason is that the CO stratification at the start of the thermal pulse phase is determined by the value 
of the He-core mass at the onset of the He-burning -- that is approximately the same in both progenitors -- not by the 
total mass. 
As a general property of the final CO profiles, the inner part of the core, 
with a constant abundance of oxygen (see Salaris et al.~1997 for more 
details on this issue) is determined by the maximum extension of the central He-burning 
convective region. Beyond this region, the oxygen profile is built when the thick He-burning shell moves toward the surface. 
During this phase, gravitational contraction increases temperature and density of the shell, and since the ratio between 
the $^{12}C(\alpha,\gamma)^{16}O$ and 3$\alpha$ reaction rates is lower for larger temperatures 
(see, e.g., Fig. 1 in Mazzitelli \& D'Antona~1987), the oxygen mass fraction steadily decreases in the external 
part of the CO core.

\subsubsection{Progenitor metallicity and IFMR}

The profiles labelled as 5 and 6 in Fig.~\ref{profiles} show the case of progenitors with metallicity equal to Z=0.002 and 
Z=0.04 (with main sequence core overshooting, as for all other calculations discussed below) respectively. 
The mass of the progenitor with Z=0.002 is $\sim1.0 M_{\odot}$ lower than our reference case, whereas it is 
approximately the same 
for the Z=0.04 model. The central value of the oxygen abundance is almost the same (within at most $\sim$0.02 in mass fraction) 
as the reference choice. The only major change is the mass extension of the inner region with the highest oxygen abundance.

Profile number 4 shows the oxygen abundance profile 
obtained using a different IFMR from the reference case. The progenitor metallicity is 
Z=0.0198 as in the reference case, but we started with a mass $\sim 1.0M_{\odot}$ lower 
($\sim 2M_{\odot}$ instead of $\sim 3M_{\odot}$) and let the model evolve through several 
thermal pulses until the core reached 0.61$M_{\odot}$. 
This choice for the progenitor is roughly consistent with the semiempirical 
IFMR determined by Salaris et al.~(2009). 

The central oxygen abundance is larger (by about 5\%) 
than the reference case, because of the smaller initial progenitor mass, that produces a 
higher central O-abundance. The extension of the central region with the highest oxygen content is smaller, 
reflecting the reduced size of both the convective inner regions, and the overall He-core during the central He-burning phase. 

\subsubsection{Breathing pulse suppression and $^{12}C(\alpha,\gamma)^{16}O$ reaction rate}

Profiles labelled as 2, 3 and 7 display the effect of two major intrinsic uncertainties in the derivation of 
evolutionary CO profiles for WDs. Profile number 2 shows the result at the first pulse for a progenitor with Z=0.0198, but 
the breathing pulses suppressed following a different method devised by Dorman 
\& Rood (1993 -- see, e.g., Cassisi et al.~2003 for more details 
on this issue). The major effect is on the central oxygen mass fraction, that increases by $\sim$15\% 
(see also Straniero et al.~2003 for a similar test). 

Profiles 3 and 7 show the abundances (for progenitor metallicity Z=0.0198) at the first pulse using 
the lower and upper limits of our adopted $^{12}C(\alpha,\gamma)^{16}O$ reaction rate, from Kunz et al.~(2002).
It is important to notice how more recent estimates of the astrophysical S factor for this reaction (Katsuma 2008, 
Dufour \& Descouvemont 2008) provide values roughly within the limits given by Kunz et al.~(2002).

The progenitor mass is essentially the same as our reference case, but the final 
profiles are obviously affected by the different rates.
We also displayed, for comparison, the oxygen profile (profile 1) used by S00 and taken 
from Salaris et al.~(1997) progenitor calculations. 
The main reason for the sizable differences with our reference choice 
is that Salaris et al.~(1997) adopted the $^{12}C(\alpha,\gamma)^{16}O$ rate by Caughlan et al~(1985).  
The more recent calculations by Kunz et al.~(2002) give a lower reaction rate, that causes an overall 
smaller oxygen abundance at the same value of the core mass.

\subsubsection{Discussion}

\begin{figure}
\epsscale{1.00}
\plotone{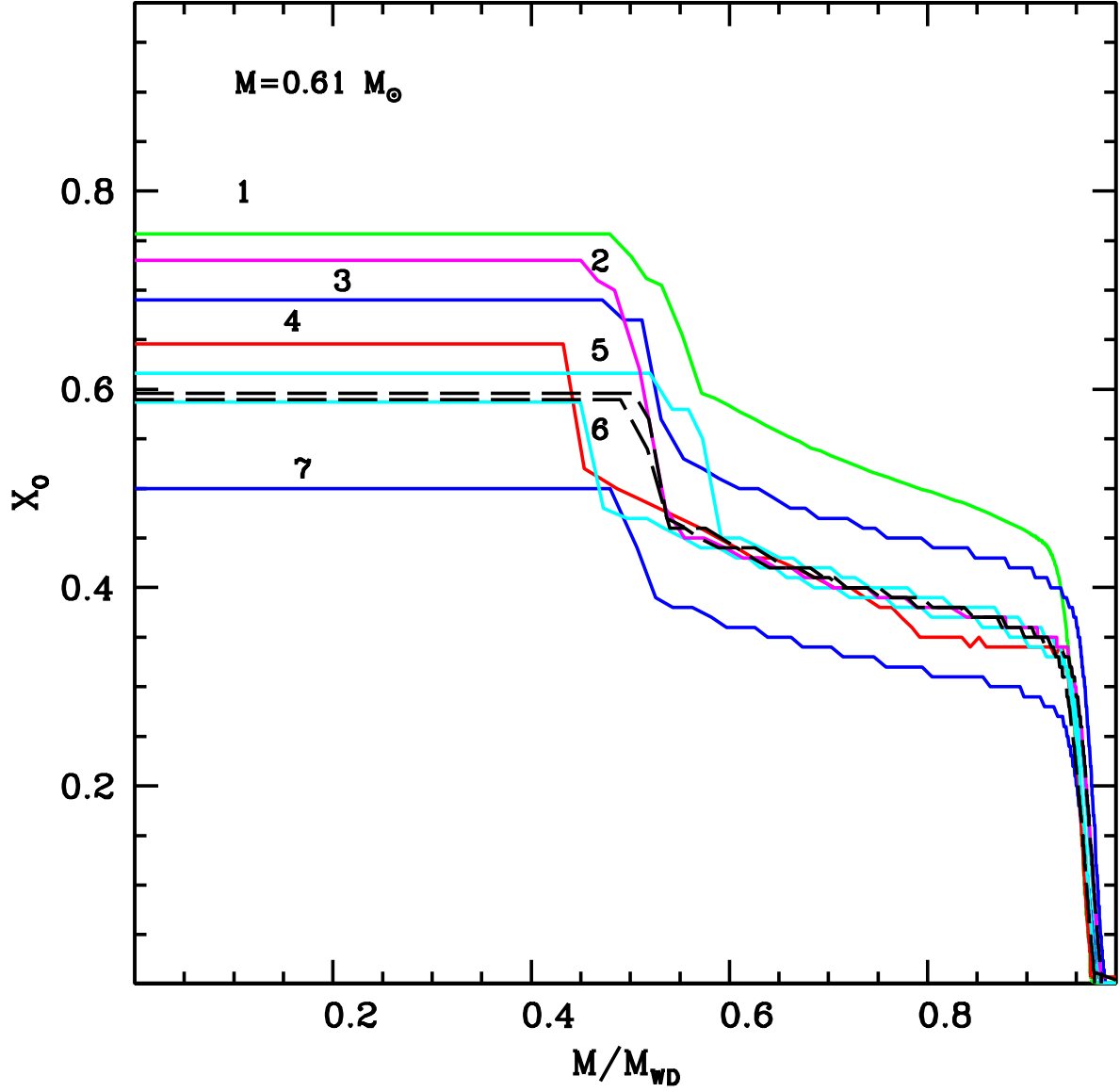}
\caption{Several different oxygen stratification (in mass fraction) tested 
on a 0.61~$M_{\odot}$ WD model. See text for details.}
\label{profiles}
\end{figure}

\begin{figure}
\epsscale{1.00}
\plotone{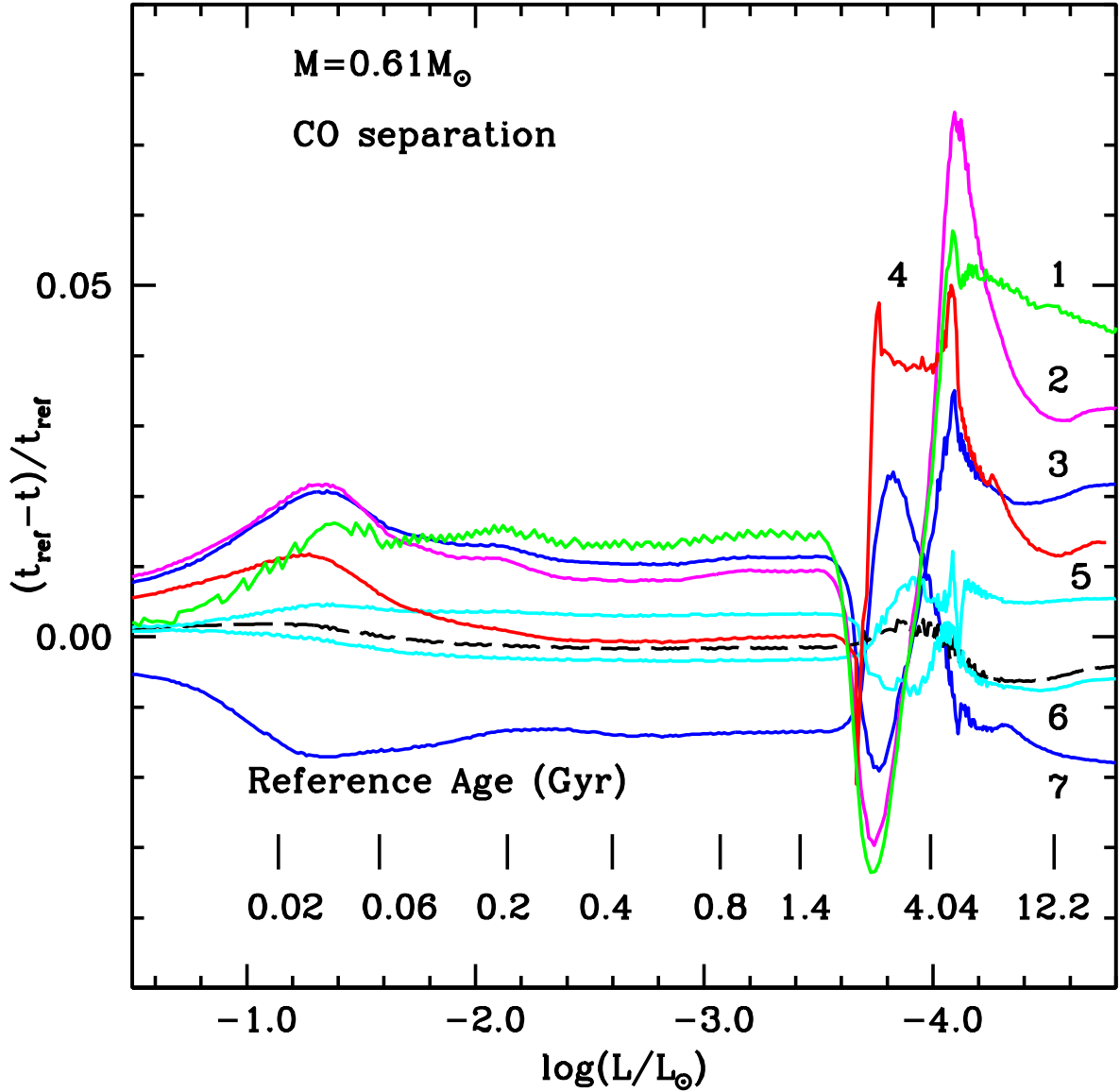}
\caption{Fractional difference between the cooling ages of models with our reference profile ($t_{ref}$)  
and models with the alternative choices displayed in Fig.~\ref{profiles} ($t$).  
Labels correspond to the profiles displayed in Fig.~\ref{profiles}. The dashed line 
denotes the difference with respect to the stratification obtained with the same assumptions of our reference choice, 
but without considering main sequence core overshooting for the progenitor evolution. Selected ages for the model 
with the reference O-profile are also displayed.}
\label{diff1}
\end{figure}

It is clear from Fig.~\ref{profiles} that the uncertainties related to the $^{12}C(\alpha,\gamma)^{16}O$ rate and 
the breathing pulse suppression 
have on the whole the largest influence on the WD chemical stratification.  
Figure~\ref{diff1} compares the cooling times of several 0.61$M_{\odot}$ H-atmosphere WD models (including the effect 
of phase separation) computed employing the profiles 
discussed above. We display the fractional difference between the cooling ages of a model with our reference profile 
and models with each one of the other choices  
shown in Fig.~\ref{profiles}. During the pre-crystallization phase 
($\log(L/L_{\odot})$ above $\sim -$3.6 and ages below $\sim$2.0~Gyr for the reference model) 
cooling times differ overall by less than $\pm$2\%, the largest variations being caused by the $^{12}C(\alpha,\gamma)^{16}O$ rate 
and breathing pulse suppression uncertainties. As a general rule, during this phase 
the models with a higher oxygen abundance tend to cool faster, as expected. 
Below $\log(L/L_{\odot}) \sim -4.2$  (ages above $\sim$ 5~Gyr for the reference model) when the crystallization of the CO core  
is almost completed, $^{12}C(\alpha,\gamma)^{16}O$ and breathing pulses are again the major sources of uncertainty, at the level of 2-3\%. 
At luminosities and ages intermediate between these two regimes the situation is more complex, because the 
exact luminosity of the 
onset of crystallization -- and the associated energy release due to latent heat and phase separation -- varies 
with changing abundance profiles. In general, the higher the central oxygen abundance, the earlier the onset of 
crystallization. This explains the 
narrow luminosity range (around $\log(L/L_{\odot}) \sim -3.6$ ) where cooling times are generally longer for models with higher 
central oxygen. During the crystallization process the exact values of the cooling times depend on the detailed shape 
of the CO profile. As a result, at luminosities between $\log(L/L_{\odot}) \sim -3.6$ and $\log(L/L_{\odot}) \sim -4.0$, the model 
with a lower mass 
progenitor that mimics a realistic IFMR  
displays the longest cooling times. Differences with our reference choice are still below 5\%. 

When the effect of phase separation upon crystallization is neglected, the qualitative behaviour of the age differences 
is exactly the same discussed before, but quantitatively the fractional differences are smaller, 
always within $\sim \pm$3\%, at all luminosities/ages.

To summarize, Fig.~\ref{diff1} shows that selecting CO profiles from progenitor 
models at the first thermal pulse, and discarding metallicity effects on the progenitor evolution, does not introduce 
major uncertainties in the cooling times 
of models at fixed $M_{WD}$. On the whole, a larger effect -- still within at most 
7\%, or 3\% when phase separation is neglected -- is caused by the treatment of convection in the 
late stages of the progenitor central He-burning phase, and the uncertainty on the $^{12}C(\alpha,\gamma)^{16}O$ rate. 
Only during the first stages  
of crystallization the choice of the IFMR has the largest impact on the cooling timescales, 
but the effect is within 5\%, and at most 3\% when phase separation is neglected.

It is also important to mention briefly the crucial role played by the thickness 
of the H-envelope. A decrease of $q_{H}$ from 
$q_{H}=10^{-4} M_{WD}$ to $q_{H}=10^{-5} M_{WD}$, keeping the core
chemical composition unchanged, 
speeds up the cooling of the models, causing a maximum age difference of $\sim$7\% at ages above $\sim$4~Gyr, in case of both 
including and neglecting the effect of phase separation.

%
\section{Cooling sequences and isochrones}
\label{models}
%

\begin{figure}
\epsscale{1.00}
\plotone{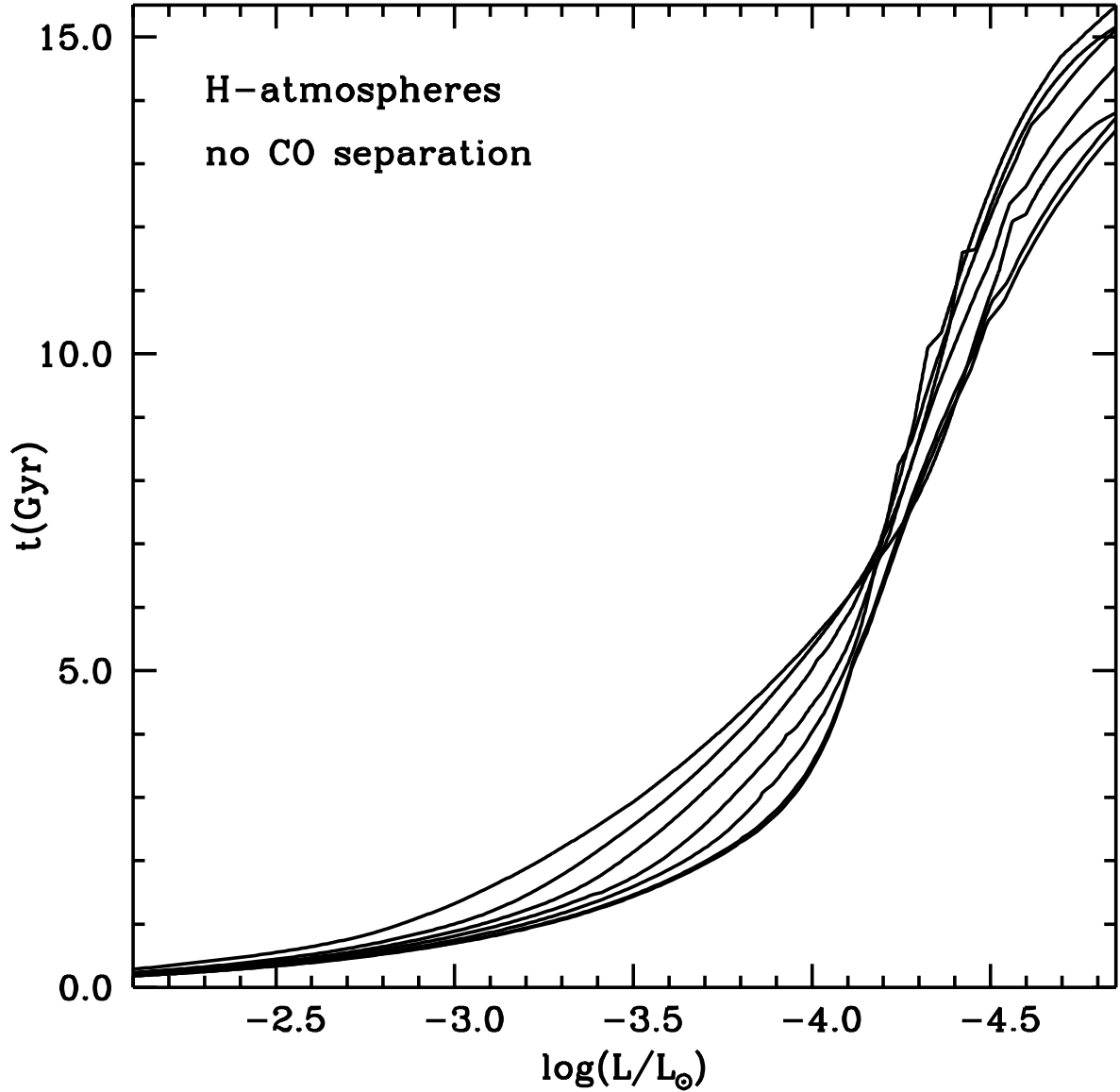}
\caption{Cooling times as a function of the surface bolometric luminosity for our complete set of 
H-atmosphere models (including crystallization but without phase separation). 
At a reference $\log(L/L_{\odot})=-3.5$, from bottom to top, 
the different lines denote the 0.54, 0.55, 0.61, 0.68, 0.77, 0.87 and 1.0 $M_{\odot}$ model, respectively.}
\label{cooling}
\end{figure}

\begin{figure}
\epsscale{1.00}
\plotone{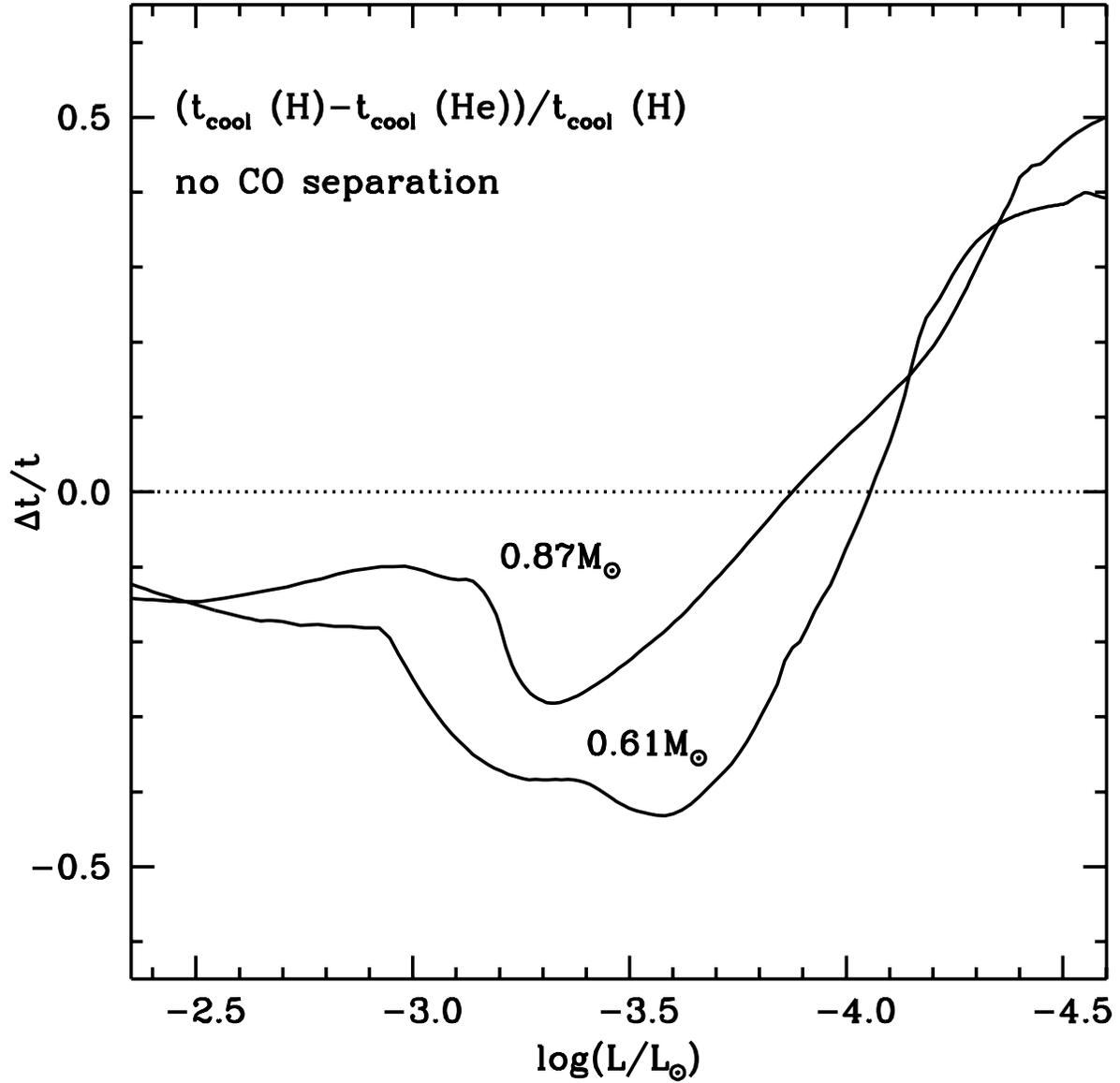}
\caption{Fractional difference between the cooling times of H- and He-atmosphere WD models 
(phase separation not included) with masses equal to 0.61 and 0.87$M_{\odot}$, respectively.}
\label{diffcool}
\end{figure}

Our new sets of WD cooling models have been computed 
with the choices for the envelope and core stratifications 
discussed in detail in the previous section. The cooling times as a function of luminosity 
for the full set of WD masses are shown in Fig.~\ref{cooling}, in case of models with H-atmospheres 
and no CO phase separation upon crystallization. The 
fractional age difference $\Delta t/t$ with respect to He-atmosphere WD calculations 
(no phase separation included) for two selected WD masses is shown in Fig.~\ref{diffcool}. 
At luminosities above $\log(L/L_{\odot})\sim -4.0$ (the exact value depending on $M_{WD}$) 
He-atmosphere models predict longer cooling times
(up to $\sim$30-50\%) due first to the higher opacity of their envelopes in this 
luminosity range (see Fig.~\ref{LTc}) and then, when their envelopes  
become less opaque, to an earlier onset of crystallization and 
earlier latent heat release.  
Below $\log(L/L_{\odot})\sim -4.0$ the H-atmosphere WDs show progressively longer cooling times, 
due to the much higher opacity of their envelopes, and $\Delta t/t$ reaches values up to 50\% at 
$\log(L/L_{\odot})\sim -4.6 - -4.7$, the faintest luminosities of our He-atmosphere calculations.

\begin{figure}
\epsscale{1.00}
\plotone{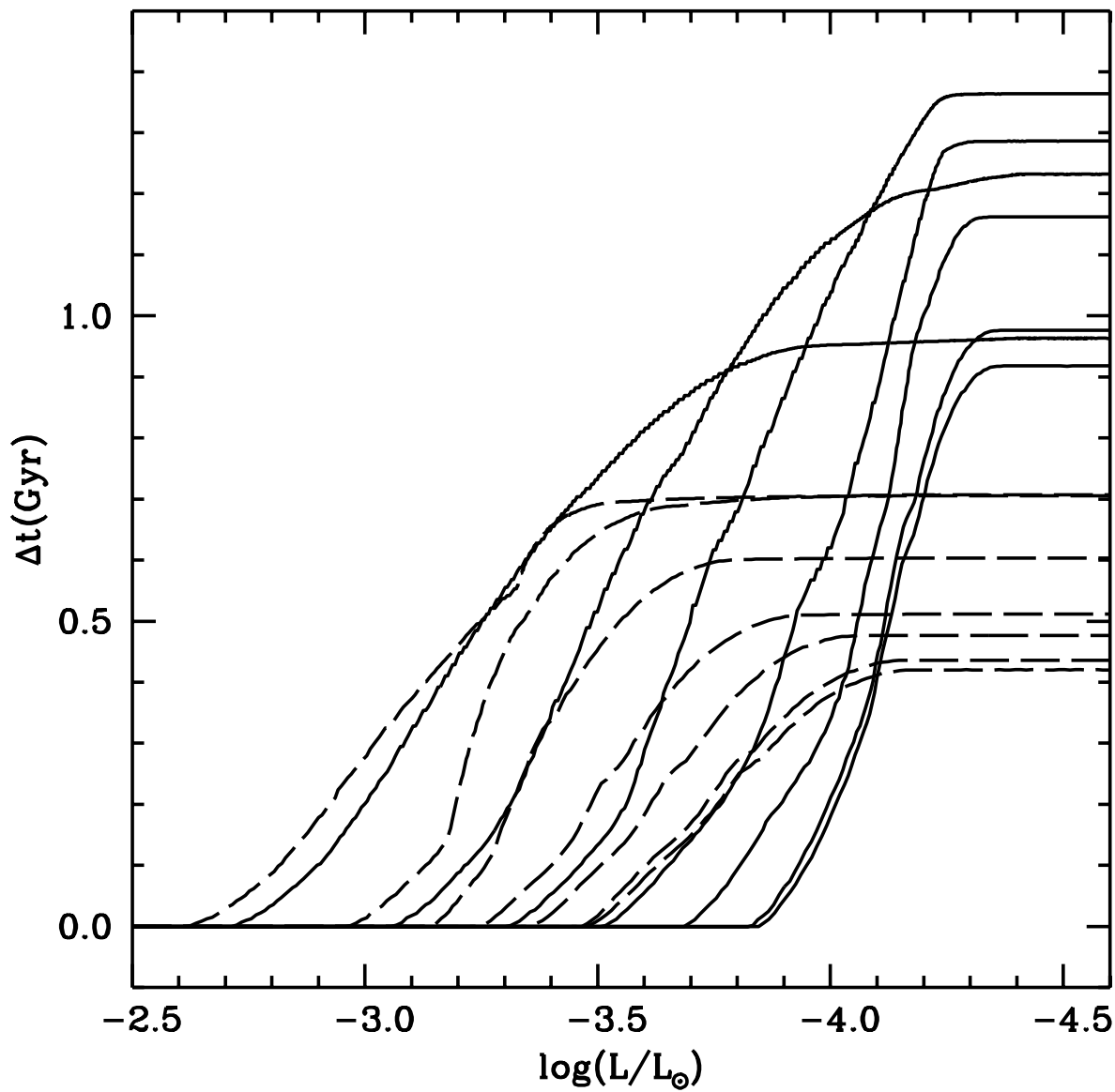}
\caption{Time delay $\Delta$t caused by the inclusion of CO phase separation 
upon crystallization, as a function of the WD luminosity. 
From right to left, the different lines denote the 
0.54, 0.55, 0.61, 0.68, 0.77, 0.87 and 1.0 $M_{\odot}$ model, respectively. 
Dashed (solid) lines represent He- (H-) atmosphere WD models. The total time delay 
$t_d$ corresponds to the final, constant value of $\Delta$t, when crystallization is completed.}
\label{delay}
\end{figure}

The total time delay $t_d$ caused by the inclusion of CO phase separation is displayed 
in Fig.~\ref{delay} for the full set of H- and He-atmosphere WD calculations. 
As in case of S00 calculations, for H-atmosphere WD models 
$t_d$ increases with mass, has a maximum at $M_{WD}$=0.77$M_{\odot}$, and  then decreases. 
Because of the different CO profile, with C/O ratio typically closer to 1 in the central regions, 
our new H-atmosphere models display on average $\sim$100~Myr larger $t_d$ at fixed $M_{WD}$, compared to 
the S00 results. In case of He-atmosphere WD calculations $t_d$ 
increases with mass and reaches a maximum for the 0.87 and 1.0$M_{\odot}$ models. 
In quantitative terms $t_d$ is roughly a factor of 2 larger for the H-atmosphere models, due essentially 
to the higher opacity of their envelopes when crystallization sets in. 

\begin{figure}
\epsscale{1.00}
\plotone{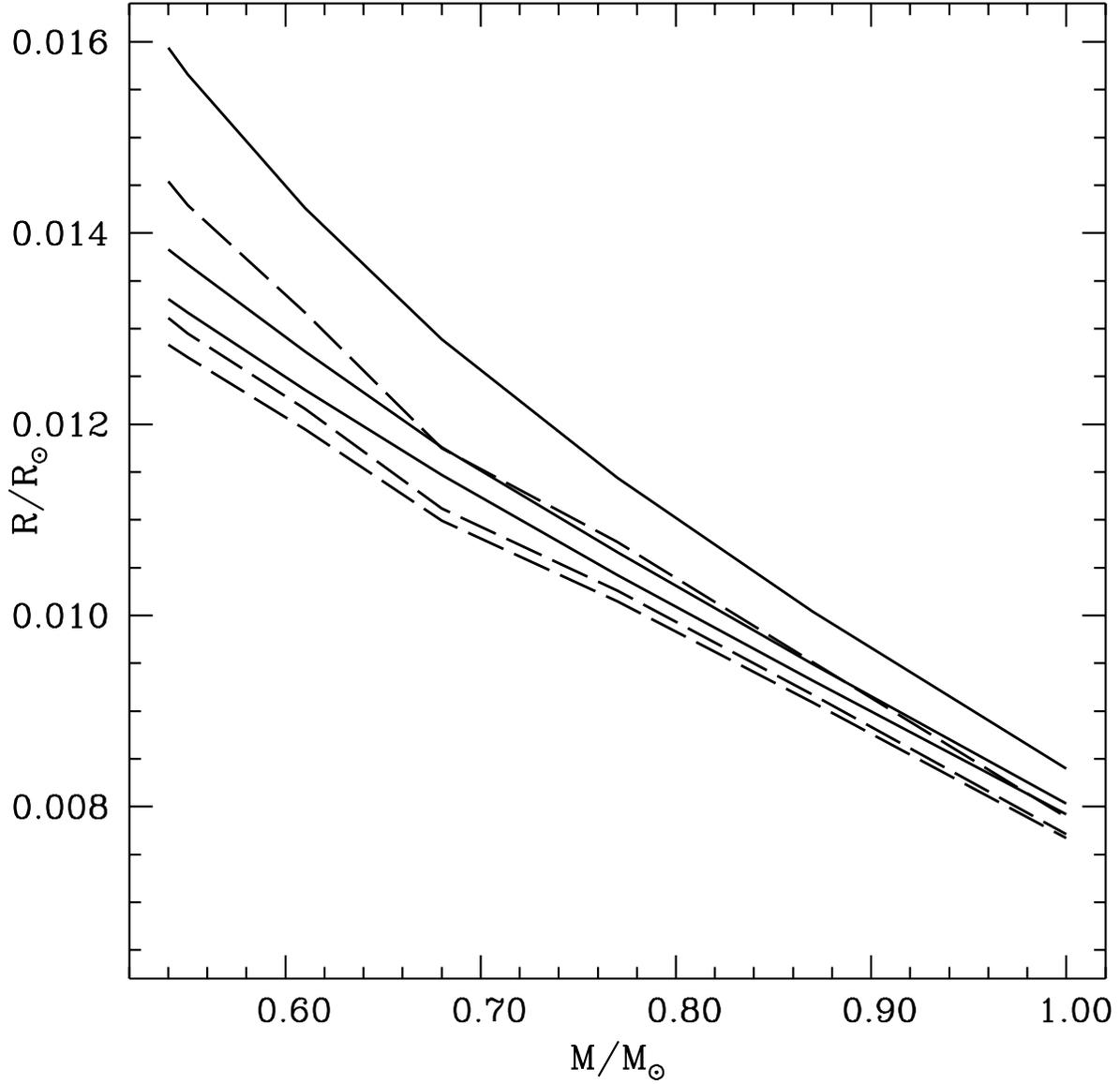}
\caption{Mass-radius relationship (in solar units) for our H- (solid lines) and He-atmosphere (dashed lines) models, 
taken at $T_{eff}$ equal to (moving from top to bottom) 30000, 10000 and 5000~K, respectively.}
\label{mradius}
\end{figure}

Figure~\ref{mradius} displays the mass-radius relationship for both H- and He-atmosphere WD models, taken at three 
different effective temperatures along the cooling sequences.
As already investigated by, e.g. Bergeron, Saffer, \& Liebert~(1992)  
the radii of He-atmosphere WDs of a given mass turn out to be systematically smaller than the H-atmosphere case.
For our models, typical differences at $T_{eff}=$30000~K are of $\sim$9\% for $M_{WD}$=0.54$M_{\odot}$, decreasing down 
to $\sim$6\% for $M_{WD}$=1.0$M_{\odot}$. When $T_{eff}=$ has decreased to 5000~K the differences are 
$\sim$4\% for $M_{WD}$=0.54$M_{\odot}$ and $\sim$3\% for $M_{WD}$=1.0$M_{\odot}$

Starting from the cooling models we have computed WD isochrones, i.e. 
the CMD of WDs born from a single-age, single-metallicity population. 
Isochrones are a 
a fundamental tool for stellar population dating, and are routinely used to study the cooling sequences 
of WDs in star clusters. Computations of WD isochrones require, in addition to   
a grid of WD models covering the relevant mass range, an IFMR, plus evolutionary timescales of the WD progenitors. 

\begin{figure}
\epsscale{1.00}
\plotone{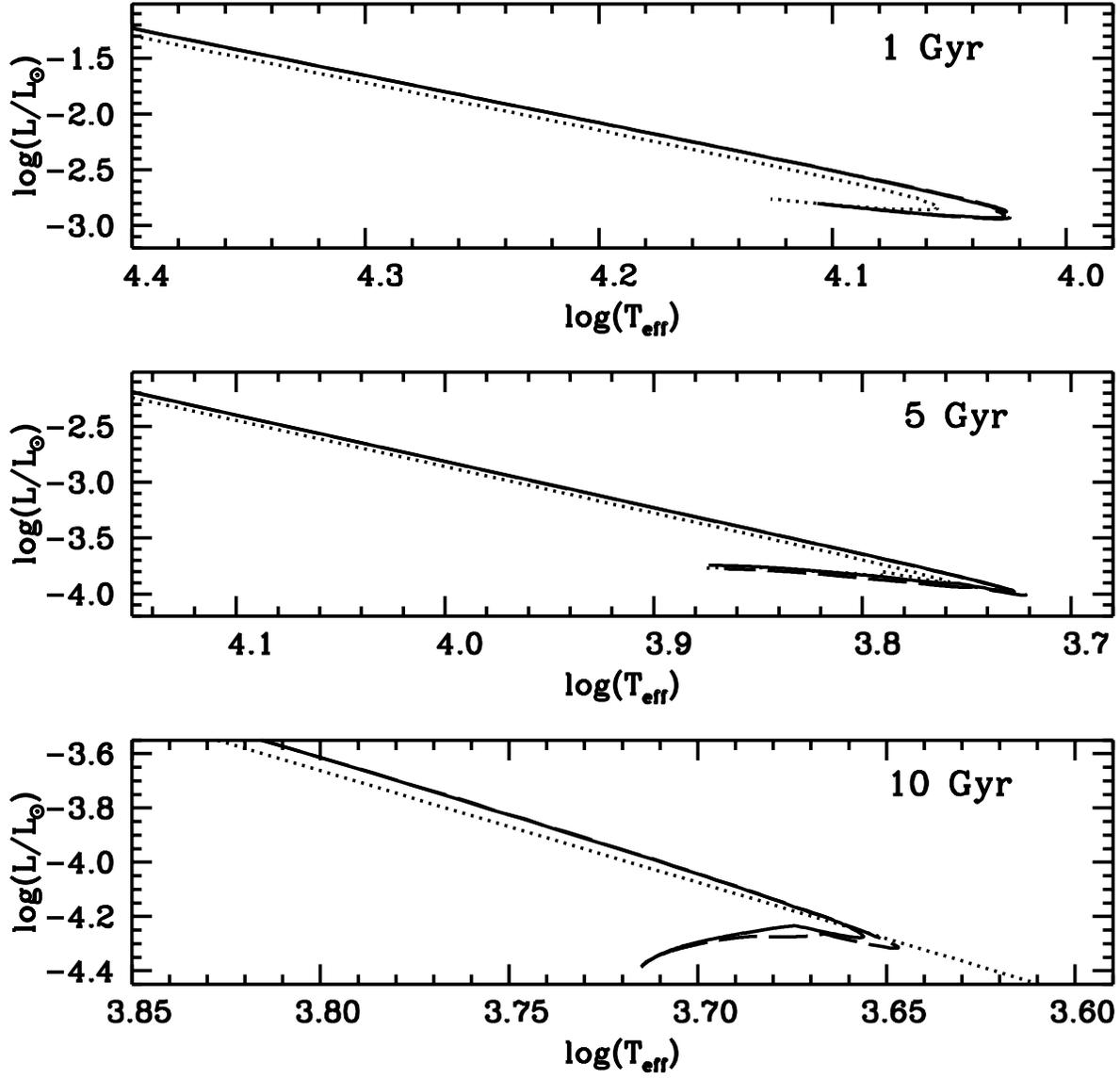}
\caption{Three sets of WD isochrones (including phase separation) 
for each of the labelled ages. Solid lines denote H-atmosphere isochrones, 
dotted lines He-atmosphere isochrones, and dashed lines 
isochrones calculated from the S00 WD models (see text for details). 
Note the change of scale between panels.}
\label{iso}
\end{figure}

Figure~\ref{iso} shows three sets of isochrones in the $L-T_{eff}$ plane, 
with ages equal to 1, 5 and 10~Gyr, for our 
H-atmosphere (solid lines) and He-atmosphere (dotted line) cooling models including phase separation upon crystallization. 
We also include (dashed lines) isochrones for the S00 models (with hydrogen atmospheres) again including phase separation.
All displayed isochrones have been computed using the IFMR by Salaris et al.~(2009)\footnote{We employed 
the linear analytic fit, extrapolated -- when necessary -- down to the smallest value of the progenitor mass appropriate for the 
chosen isochrone age. The upper value of the initial mass is set by 
the minimum stellar mass igniting carbon core burning, as derived from the BaSTI calculations}  
and progenitor lifetimes from BaSTI models (with convective core overshooting during the main sequence) 
for a metallicity Z=0.0198. 
As discussed before, the choice of the progenitor metallicity and IFMR has a small impact on the CO stratification 
and cooling times of models with fixed $M_{WD}$, and one can couple our sets of WD models to different choices for  
the IFMR and progenitor metallicity.

As is well known (see, e.g. Salaris~2009 for a review), the age indicator is the faint end of the isochrones. 
The more massive WDs formed from higher-mass and shorter-lived progenitors pile up at the bottom of 
the cooling sequence, where they produce the characteristic turn to the blue, i.e. a turn towards lower radii 
(Isern et al.~1998). An  
age increase makes the bottom end of the isochrones fainter, because of the longer cooling times. 

A qualitative analysis of Fig.~\ref{iso} shows that H-atmosphere WD isochrones computed from S00 models, 
have an only slightly fainter termination at ages of 5 and 10~Gyr, compared to our results.
On the other hand our He-atmosphere WD isochrones have a brighter faint end than the H-atmosphere counterpart  
at ages of 1 and 5~Gyr, while at 10~Gyr they reach much lower luminosities. 
Given that progenitor ages and IFMR are the same for all three sets of isochrones, this behavior is due to differences in the 
cooling times of the underlying WD models. As we have seen before, for luminosities down to 
$\log(L/L_{\odot})\approx-$4, He-atmosphere models predict a slower cooling, 
hence the brighter termination of the WD isochrones for 1 and 5~Gyr. 

Another interesting result of the comparisons in Fig.~\ref{iso} is the general shift 
of He-atmosphere isochrones towards lower radii, very evident when the luminosity is above $\log(L/L_{\odot})\approx-$4. 
There are two reasons for this behavior. The first one is related to the fact that 
along a WD isochrone of age $t$, the sum of the 
WD cooling age and the corresponding progenitor lifetime has to be equal to $t$. 
Above $\log(L/L_{\odot})\approx-$4 He-atmosphere models with a given 
$M_{WD}$ have longer cooling times and, as a consequence, a given luminosity along an isochrone has to be populated by 
a larger WD mass 
(smaller radius) in the non-DA case, because its earlier formation (lower progenitor lifetimes) compensates for the 
longer cooling times.
An additional contribution to this difference stems from the fact that, as shown in Fig.~\ref{mradius}, 
He-atmosphere WD models of a given $M_{WD}$ have (at fixed $T_{eff}$) smaller radii than the H-atmosphere counterpart.

\begin{figure}
\epsscale{1.00}
\plotone{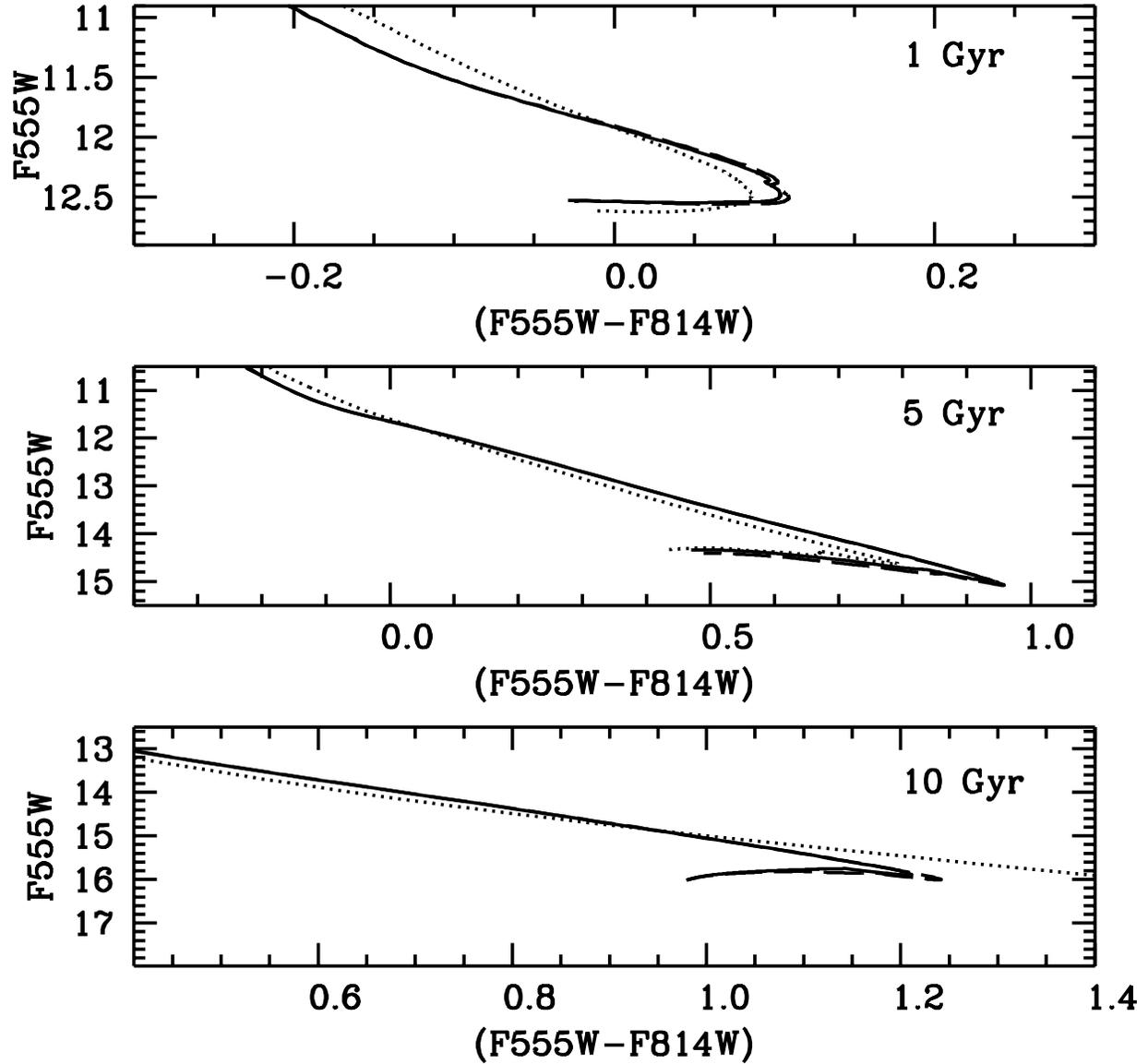}
\caption{ The same isochrones of Fig.~\ref{iso}, this time in the ACS F555W-(F555W-F814W) plane. 
The line-styles are as in Fig.~\ref{iso}.}
\label{isoACS}
\end{figure}

\begin{figure}
\epsscale{1.00}
\plotone{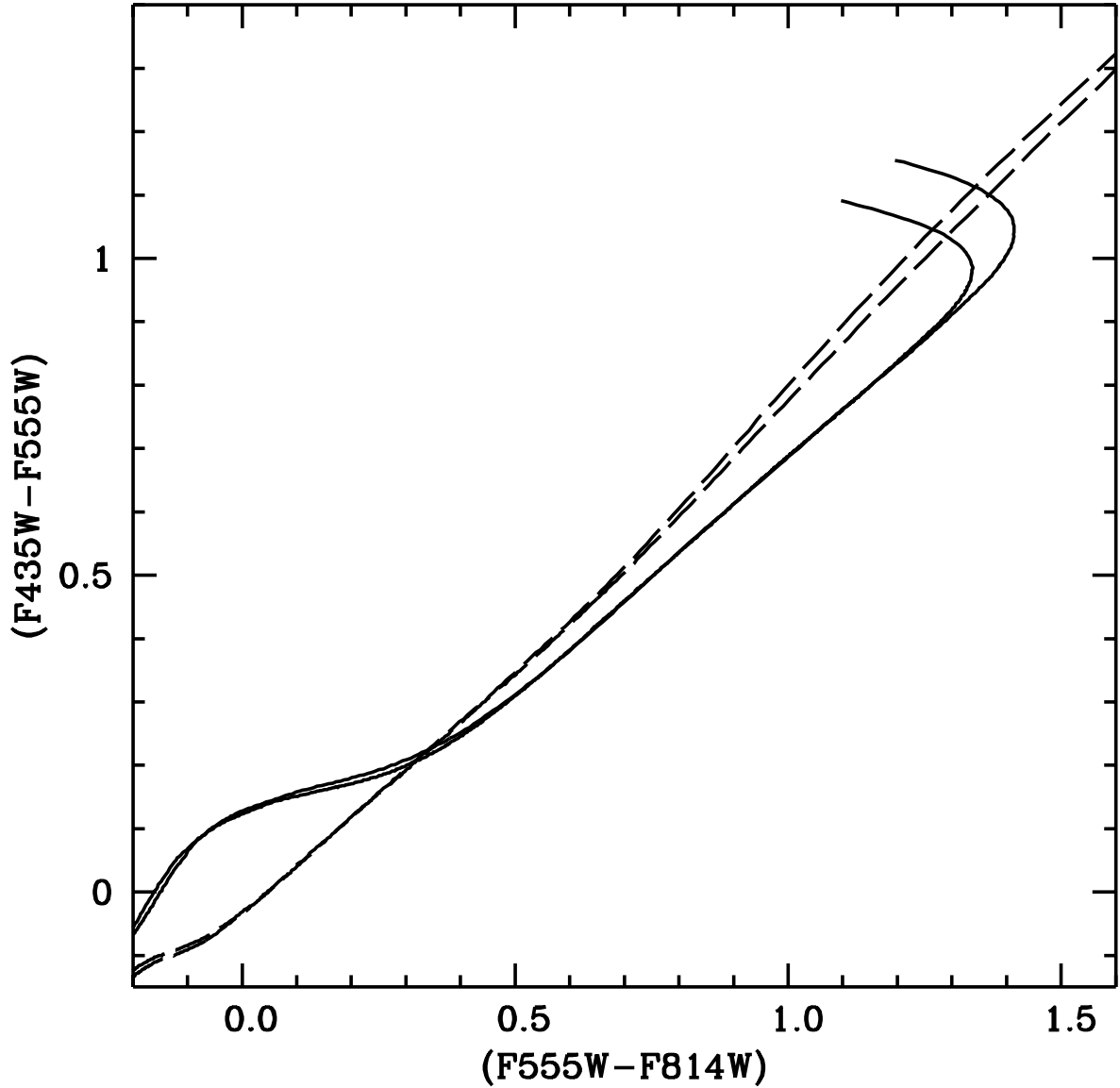}
\caption{ (F435W-F555W) vs (F555W-F814W) diagram 
in the ACS photometric system, for the 0.61 and 1.0$M_{\odot}$ cooling tracks (including phase separation) 
with H- (solid lines) and He-atmospheres (dashed lines).}
\label{colors}
\end{figure}

Figure~\ref{isoACS} displays the same sets of isochrones, this time in an observational plane, i.e. employing 
the absolute magnitudes in the F555W and F814W filters of the ACS camera on board HST. 
The reference set of bolometric corrections (used for Fig.~\ref{isoACS}) that we apply to all our WD models 
and isochrones is the same employed in Bedin et al.~(2005), i.e. an extension of 
the results of Bergeron, Wesemael \& Beauchamp (1995 - see, e.g. Sect.3.1 in Holberg \& Bergeron~2006). 
One can notice how the different 
behavior of the bolometric corrections for H- and He-atmospheres alters the relative location 
of the corresponding isochrones, 
compared to Fig.~\ref{iso}. Despite the longer cooling times of He-atmosphere WD models, the termination of the 
isochrone at 1~Gyr is now fainter than the H-atmosphere one.
The differences in the bolometric corrections and colors of our H- and He-atmosphere cooling tracks can be appreciated even better in 
Fig.~\ref{colors}, that displays a color-color diagram for the 0.61 and 1.0$M_{\odot}$ tracks with both H 
(solid lines) and He atmospheres 
(dashed lines). The He-atmosphere colors increase steadily along the evolution, whereas the (F555W-F814W) color 
of the H-atmosphere models display a more complex behaviour, with a marked decrease at low effective temperatures,  
due to the blocking effect in the infrared of the H$_2$ collision-induced absorption (e.g., H99, Saumon \& Jacobson~1999). The onset 
of this turn to the blue of (F555W-F814W) is at ages above 14~Gyr, but is attained earlier (and the turn to the blue more pronounced) 
in near-infrared colors.
In the future we will update the adopted set of bolometric corrections by calculating theoretical spectra 
from the new atmospheres employed for the boundary conditions of our WD models. 
In the new pure H model atmospheres, the updated calculation of the $Ly_{\alpha}$ opacity 
causes the removal of flux from short
wavelenghts and its redistribution to longer wavelengths (Kowalski~2007), compared to previous calculations. 
As for the new pure He 
model atmospheres, a stronger ionization makes the He-ff opacity dominant
over Rayleigh scattering at all wavelenghts (Kowalski et al.~2007), and the spectral energy distribution 
is closer to a black body compared to Bergeron, Wesemael \& Beauchamp~(1995) results.
Some preliminary estimates for gravities typical of a 0.6$M_{\odot}$ WD model, 
show that color transformations from the new model atmospheres will produce redder 
colors for both H- and He-atmospheres.

One can derive a quantitative estimate of 
the age differences obtained when the three sets of isochrones in Fig.~\ref{isoACS} are applied to real data, 
in the following way. 
We have first calculated the LFs in the F606W passband for the three H-atmosphere isochrones 
in  Fig.~\ref{isoACS}, assuming a Salpeter mass function (MF) for the progenitors. 
For the three reference ages of 1, 5 and 10~Gyr, we have determined the 
magnitude of the LF cut-off, that corresponds to the faint end of the isochrones. 
These three LFs are considered to be the 'observed' LFs of three populations of known ages. 
We have then computed 
several isochrones and LFs from both our He-atmosphere calculations, and from the older S00 models, 
and determined what ages are necessary to match the position of the LF cut-offs of the three reference 'observed' 
populations.


The results, reported in Table~1, show that S00 models provide ages very similar to our new H-atmosphere cooling models, 
across the whole range explored by this test. This implies that all WD ages in the 
series of papers by Bedin and collaborators (e.g. Bedin et al.~2010 and 
references therein) obtained using S00 models (and BaSTI progenitor lifetimes) 
are basically confirmed by our new calculations. Also the derivation of the 
semiempirical IFMR by Salaris et al.~(2009), that makes use of 
H-atmosphere WD ages estimated from S00 models, is basically unaffected.
On the other hand He-atmosphere isochrones give, as well known, 
much younger ages for the oldest population, but 10\% older ages at 5~Gyr, 
and 10\% younger ages at 1~Gyr. 

\begin{table}
  \begin{center}
  \label{tab1}
  \caption{Correspondence between the LF cut-off ages for the following three sets of models (see text for details).}
 {
  \begin{tabular}{|l|l|l|l|l|}\hline 
{\bf H-atm. } & {\bf He-atm. } & {\bf H-atm. S00 } \\\hline
1.0~Gyr & 0.9~Gyr & 1.0~Gyr \\\hline
5.0~Gyr & 5.5~Gyr & 4.7~Gyr \\\hline
10.0~Gyr & 6.5~Gyr & 9.7~Gyr \\\hline
  \end{tabular}
  }
 \end{center}
\end{table}

\subsection{The WD population in the solar neighborhood}

\begin{figure}
\epsscale{1.00}
\plotone{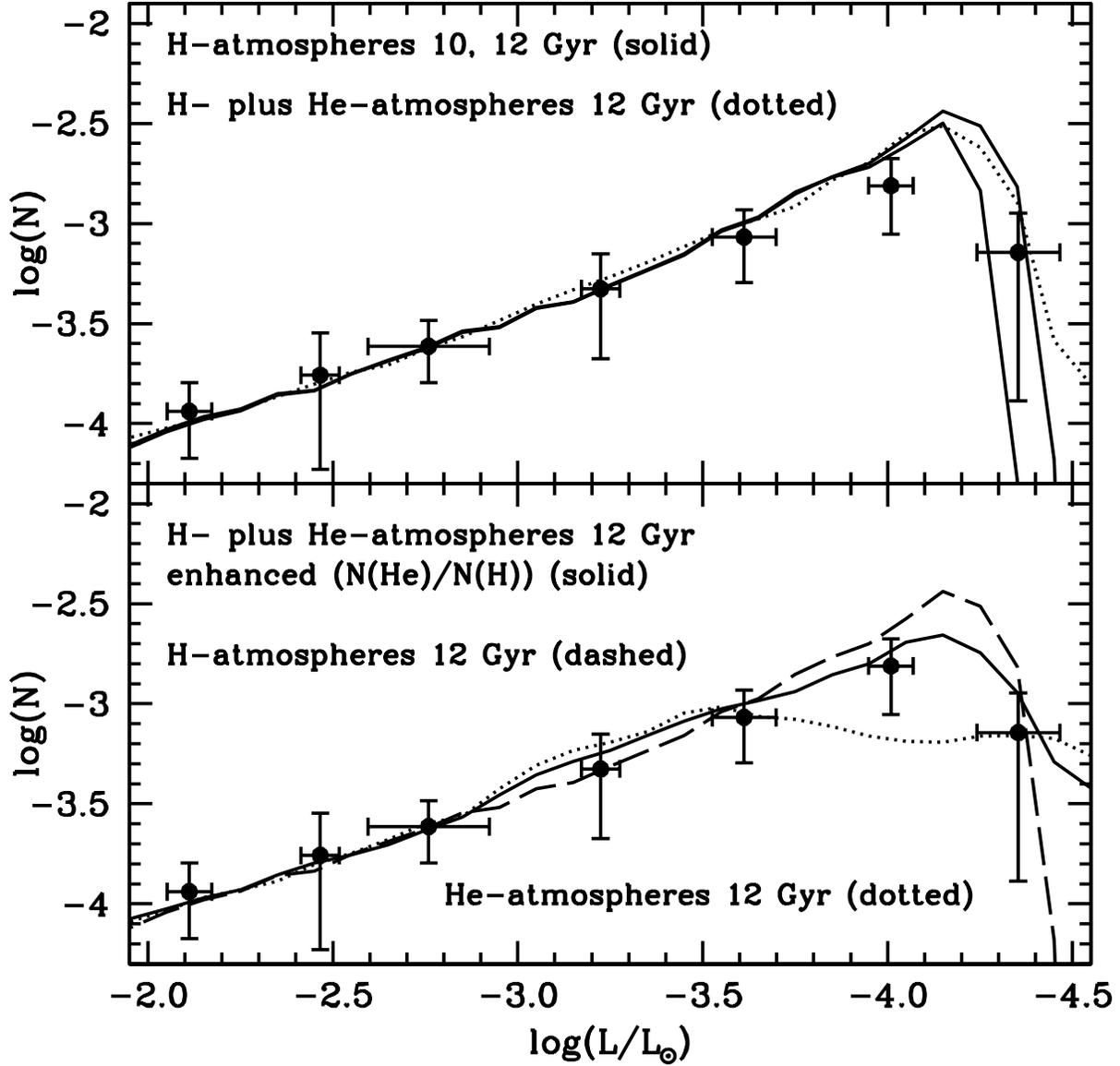}
\caption{Upper pannel: Observed LF for WDs in the solar neighborhood from Catalan et al.~(2008) compared, respectively, to 
H-atmosphere theoretical LFs and a mixed H- and He-atmosphere LF with a number 
ratio (N(He)/N(H)) equal to the observed mean value at $T_{eff}$=14000~K  
(see text for details). Lower panel: The observed LF is compared to, respectively, H- and He-atmosphere LFs, and 
a LF with an enhanced (N(He)/N(H)) ratio (see text for details).
All theoretical LFs are computed from WD models including the effect of phase separation upon crystallization.}
\label{Disk}
\end{figure}

We close this section with an example of application of our models to real data.
Figure~\ref{Disk} displays an observational LF of WDs in the solar neighborhood,  
from Catalan et al.~(2008  -- compiled from several sources) compared to several theoretical LFs. 
More in detail, we calculated LFs from both our H-atmosphere (with and without phase separation) 
and He-atmosphere (only the ones with phase separation included) models, considering  
progenitors with metallicity Z=0.0198 (including main sequence convective core overshooting) 
and a Salpeter MF, plus IFMR from Salaris et al.~(2009),   
and a constant star formation rate starting $t$~Gyr ago. 
All theoretical LFs are normalized to the observed star counts at $\log(L/L_{\odot}) = -2.76$.
Matching the position of the cut-off of this empirical LF with H-atmosphere models   
provides an age $t\sim$ 12~Gyr for the onset of star formation when phase separation is included, and $\sim$11~Gyr   
when phase separation is neglected. The formal 
error bar on $t$ is set by the horizontal error bar on the last point of the empirical LF, and is 
of the order of $\pm$2~Gyr.
We have also considered the effect of He-atmosphere WDs in the theoretical LF, by computing first 
a LF assuming the same parameters as for the H-atmosphere case. 
As a second step we have built a composite LF adding up 
the star counts in the H- and H-atmosphere LFs, whereby the number ratio (N(He)/N(H)) of He- to H-atmosphere WDs  
has been set to reproduce the observed mean value (N(He)/N(H))=0.268 at $T_{eff}$= 14000~K (Tremblay \& Bergeron~2008). 
This composite LF has been then normalized to match  
the observed WD counts at $log(L/L_{\odot}) = -2.76$.  
Figure~\ref{Disk} shows that our composite LF displays, at a given $t$, a sharp drop in star counts at 
approximately the same luminosity  
of the cut-off in the H-atmosphere LF, with a spread of objects distributed towards lower luminosities, due to the 
faster cooling times of the oldest He-atmosphere objects.
As a further test, we have computed a 12~Gyr LF with both H- and He-atmosphere objects -- constructed as described before --  
with a constant star formation rate and a progenitor metallicity equal to 
Z=0.004 for WD ages $t$ between 12 and 8 Gyr, Z=0.008 when $t$ is between 8 and 4~Gyr, and up to Z=0.0198 for 
ages below 4~Gyr. The result is barely different from the case of constant progenitor metallicity.

As mentioned before, in the composite LF with both H- and He-atmosphere objects we have normalized 
the ratio (N(He)/N(H)) by matching the observed mean value at  $T_{eff}$= 14000~K. 
Tremblay \& Bergeron~(2008) investigation shows that the observed (N(He)/N(H)) ratio increases up to $\sim$0.45 when 
$T_{eff} <$10000~K, and the authors conclude that the only physical mechanism able to account for this increase 
is the convective mixing of the thin hydrogen layers with the underlying helium envelope.
Here we study how (N(He)/N(H)) changes in our modelling of the local WDs, due exclusively to the different cooling times 
of H- and He-atmosphere models. We are assuming in this analysis that both 
types of WDs are born independently  
with the same IFMR, from progenitors formed with a constant formation rate and a Salpeter MF. Our choice of 
thick H-layers prevents any mixing between the H-rich envelope with the underlying, much more massive He-layers. 
Figure~\ref{Disk} displays two theoretical LFs, one with only H-atmosphere objects, and one with only  
He-atmosphere WDs, both normalized to the observed star counts 
at $\log(L/L_{\odot}) = -2.76$. In this way, just by comparing the two LFs,  
we have a first visual impression of the intrinsic, appreciable 
variation of the ratio (N(He)/N(H)) with luminosity  -- hence $T_{eff}$ --  
due to the different cooling timescales of the models. It is immediately clear that (N(He)/N(H)) does not stay constant 
along the LF; 
there is  
a luminosity interval, between log($L/L_{\odot}$)$\sim-$2.85 and $\sim -$3.5, where (N(He)/N(H)) increases, before dropping 
fast at lower luminosities, and eventually increasing again around the cut-off luminosity, due to the disappearance 
of H-atmosphere objects.

\begin{figure}
\epsscale{1.00}
\plotone{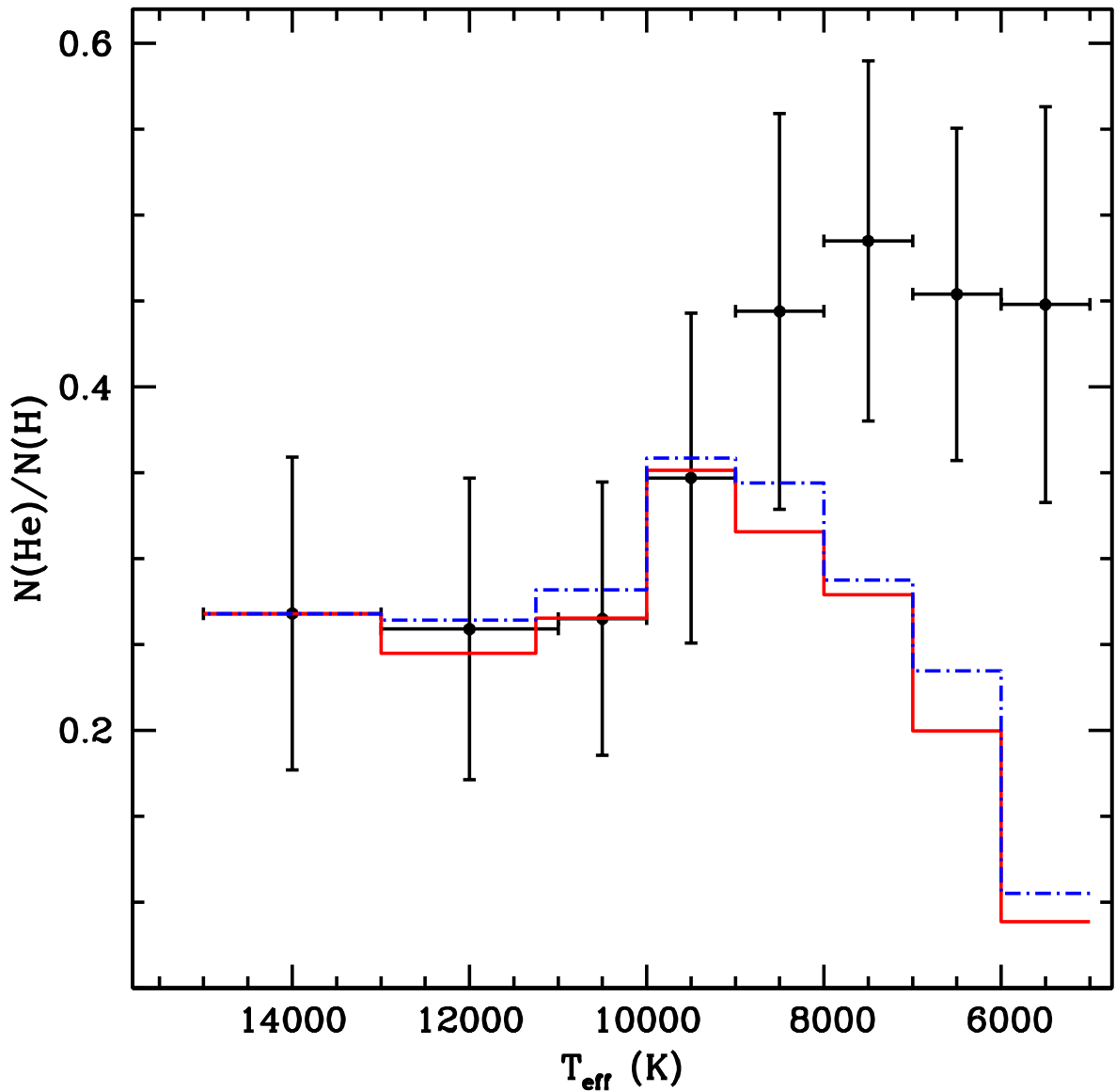}
\caption{The (N(He)/N(H)) number ratio as a function of $T_{eff}$. Points with error bars represent the 
data by Tremblay \& Bergeron~(2008). The solid line displays the predictions from the 12~Gyr theoretical LFs  
with both H- and He-atmosphere objects, shown in Fig.~\ref{Disk}. The (N(He)/N(H)) ratio is set to the observed 
mean value at 
$T_{eff}$=14000~K. The dash-dotted lines shows the predicted (N(He)/N(H)) ratio for our simulation with 
progenitor metallicity varying with age.}
\label{DAnDAratio}
\end{figure}

Figure~\ref{DAnDAratio} displays the predicted (N(He)/N(H)) number ratio (solid line) this time  
as a function of $T_{eff}$, that can be compared directly with the empirical result by Tremblay \& Bergeron~(2008).  
In this figure 
the theoretical values have been determined by means of a Monte Carlo simulation that uses as input 
the constant star formation rate, 
constant Z=0.0198 progenitor metallicity, the same IFMR and MF employed in the calculation of the LF, 
and a Galactic disk age of 12~Gyr. 
We calculate two synthetic samples of H- and He-atmosphere objects, respectively. 
For each synthetic WD produced in our simulation, we perturbed the $T_{eff}$ by a 1$\sigma$ Gaussian error equal to 5\% 
of the actual value of $T_{eff}$, to mimic the typical errors in the empirical $T_{eff}$ by Tremblay \& Bergeron~(2008). 
We have then grouped the resulting sample (over 100000 H- and He-atmosphere objects, 
to avoid statistical number fluctuations in the synthetic sample) in the same $T_{eff}$ bins 
chosen by Tremblay \& Bergeron~(2008). 
The (N(He)/N(H)) values have been first normalized to reproduce the observed mean value at 
$T_{eff}$=14000$\pm$1000 ~K, and then compared with the empirical data. 

One can notice that (N(He)/N(H)) stays roughly constant between $T_{eff}\sim$14000~K 
and $\sim$10000~K, as observed. Below this temperature the ratio increases at first, 
following the observations. This is at odds with the interpretation by Tremblay \& Bergeron~(2008); 
according to our modelling 
of the solar neighbourhood WDs, this increase is simply due to the different cooling times of H- and He-atmosphere WDs in 
this $T_{eff}$ range.
However, the theoretical value drops below the data when $T_{eff}\sim$8000~K. 
We have also determined the evolution of (N(He)/N(H)) with $T_{eff}$ in our simulation with metallicity 
increasing with decreasing WD age, and the result is not changed significantly. 
The difference between observed and predicted ratio gets larger than the 2$\sigma$ errors 
for the two coolest bins centered at 6500~K and 5500~K, respectively.
The coolest temperature bin for the observed sample corresponds approximately to the luminosity of the peaks of the theoretical 
LFs displayed in the top panel of Fig.~\ref{Disk} (log($L/L_{\odot}$)$\sim-$4.2).
For heuristic purposes it is important to mention that  
the theoretical (N(He)/N(H)) ratio reaches a minimum value of $\sim$0.06 at $T_{eff}\sim 5000$~K, beyond 
the lowest temperature limit of Tremblay \& Bergeron~(2008) data, before starting to increase. At $T_{eff}\sim 4300$~K 
the predicted value of (N(He)/N(H)) is again equal to 0.268, and increases steadily at lower temperatures, so that 
He-atmosphere objects are expected to dominate the population of the fainter bin of the observed 
LF, as is also clear from the lower panel of Fig.~\ref{Disk}.

The comparison in Fig.~\ref{DAnDAratio} shows that it still seems 
necessary to invoke the transformation of some H-atmosphere WDs into He-atmosphere objects to 
reproduce the spectroscopic observations at low $T_{eff}$, in the assumption of a constant progenitor formation rate and 
a (N(He)/N(H)) ratio at the start of the WD phase that is constant with time. 
The onset of this spectral transformation and 
the quantitative details are however different from the conclusions by Tremblay \& Bergeron~(2008), who assumed a constant baseline 
value of (N(He)/N(H)), instead of an intrinsic variation with $T_{eff}$ due to WD evolutionary effects.
The temperature where the theoretical ratio drops 
significantly below the data constrains the thickness of the H-layers in the objects that undergo the spectral 
transformation. The lower this temperature, the thicker (in mass) the H-layers. From the results in Fig.~1 of 
Tremblay \& Bergeron~(2008) and our own models, 
for the case of a 0.6$M_{\odot}$ WD, H-He mixing at $T_{eff}$=7000~K implies 
log($q(H)$)$\sim -$8.5, while mixing at $T_{eff}$=6000~K corresponds to log($q(H)$)$\sim -$7.0.
Assuming a constant observed mean value (N(He)/N(H))=0.45 when $T_{eff}$ is below 8000~K, one needs a fraction 
of H-atmosphere objects undergoing spectral transformation that  
increases with decreasing $T_{eff}$, reaching a maximum of $\sim$24\% at the lowest temperature bin sampled by 
Tremblay \& Bergeron~(2008). This is consistent 
with a broad range of H-layer thickness in solar neighborhood H-atmosphere WDs, progressively thicker H-envelopes 
being mixed at increasingly lower $T_{eff}$. 

Finally, to gain a very approximate idea of the impact of this spectral transformation on the theoretical LF, we display in 
Fig.~\ref{Disk} also the case of a mixed H- and He-atmosphere population, where (N(He)/N(H)) has been normalized appropriately to 
reach the value (N(He)/N(H))=0.45 at log($L/L_{\odot})=-$4.2 (i.e., it is about 4 times 
larger at $T_{eff}\sim$14000~K than our reference case displayed in the top panel of the same figure). 
The LF cutoff is less sharp, the mean age of the onset of star formation in the solar neighborhood is decreased by $\sim$1~Gyr.

%
\section{Summary}
\label{summary}

We have expanded our BaSTI stellar evolution archive by including new, updated WD cooling models, computed using the 
CO stratification obtained from BaSTI AGB progenitor calculations.
Improvements with respect to the S00 set of WD models concern the CO chemical profiles, that 
have been obtained employing an updated 
estimate of the $^{12}C(\alpha,\gamma)^{16}O$ reaction rate, and the inclusion of a full set of 
He-atmosphere WD models, 
computed with appropriate boundary conditions from non-gray model atmospheres.
The reference set of WD models that will be made public at the BaSTI website makes use of the CO stratification 
at the first thermal pulse from progenitor models calculated with intial metal mass fraction Z=0.0198, 
and the inclusion of convective core overshooting during 
the main sequence. To assess how sensitive the models are to these assumptions, 
we have tested the effect of uncertainties on the recent determination of  
the $^{12}C(\alpha,\gamma)^{16}O$ reaction rate employed in the progenitor models, the inclusion/exclusion of core 
convective overshooting during 
the main sequence, different approaches for quenching the breathing pulses 
at the end of core He-burning, a variation of the metallicity 
of the progenitor, a variation of the number of pulses experienced by the progenitor models. 

The results of this analysis indicate that the uncertainty on the $^{12}C(\alpha,\gamma)^{16}O$ reaction rate and 
the numerical approach 
used for inhibiting the breathing pulses have on the whole the largest impact on the WD cooling times, 
of about 7\% at most- or of about 3\% when the effect of phase separation upon crystallization is neglected.
The progenitor metallicity, convective core overshooting during the main sequence phase, number of pulses 
before the WD formation, have overall a smaller effect. 

We have discussed quantitatively differences in the mass-radius relationships and 
cooling speed of H- and He-atmosphere cooling models. 
The radii of the He-atmosphere models of a given mass are systematically lower than their 
H-atmosphere counterparts. Differences range between $\sim$ 9\% and $\sim$ 3\%, increasing with decreasing $M_{WD}$ 
and/or increasing temperature. He-atmosphere models  
show typically longer cooling times down to log($L/L_{\odot}$)$\approx -$4, before starting to cool down 
much faster at lower luminosities.
We have also estimated the differences between ages of star clusters obtained employing our new 
H- and He-atmosphere WD models, as well as the S00 H-atmosphere WD calculations. 
Ages derived from S00 H-atmosphere models show 
only relatively small differences when compared to our new calculations. 

As an example of application of our new set of models to real data, we have estimated an age of $\sim$ 12~Gyr for the 
onset of star formation in the solar neighborhood, by fitting the local WD LF compiled by Catalan et al.~(2008). 
We have also studied 
the variation of the number ratio (N(He)/N(H)) with $T_{eff}$, predicted by our simulation of the local WDs. 
Due to the different 
cooling times of H- and He-atmosphere models, we show how this ratio changes with $T_{eff}$, 
increasing below $T_{eff}\sim$10000~K, as observed. 
However, at least with our assumptions about the formation of the local WDs 
-- a constant progenitor formation rate and 
a (N(He)/N(H)) ratio at the onset of the WD phase that is constant with time --
the predicted ratio drops well below the 
observed value when $T_{eff}$ is lower than 7000-8000~K. This result can be explained in terms of 
the spectral transformation of a fraction 
of H-atmosphere objects, that increases with decreasing $T_{eff}$ below 7000-8000~K, 
reaching a maximum of $\sim$24\% at the lowest temperatures sampled by the observational data. 
As a consequence, one needs a broad range of H-layer thickness in solar neighborhood H-atmosphere WDs to explain these spectral 
changes, thicker envelopes being mixed with the underlying more massive He-layers at increasingly lower $T_{eff}$. 

All cooling tracks and the reference chemical stratifications  
will be made publicly available at the official BaSTI website 
(\url{http://www.oa-teramo.inaf.it/BASTI}). In addition, we provide WD isochrones for ages between 200~Myr and 14~Gyr for both 
H- and He-atmosphere objects (with and without the inclusion of phase separation) 
using as a reference the IFMR by Salaris et al.~(2009) and the 
progenitor lifetimes from BaSTI models including 
convective core overshooting on the main sequence. The isochrones will be available for progenitors with 
both scaled solar and $\alpha$-enhanced mixtures, 
and 11 values of the metal fraction Z, ranging from Z=0.0001 to Z=0.04. 
For both cooling tracks and isochrones we provide magnitudes in the UBVRIJHK, and HST ACS photometric systems.

%

\acknowledgments

We are deeply indebted to Didier Saumon for several invaluable suggestions  
during the whole development of this project, as well as comments on 
a preliminary version of the manuscript. 
We thank our referee for very insightful comments that helped to 
improve the presentation of our results.
J.I. acknowledges the financial support of the MICINN program AYA08-1839/ESP and
the 2009SGR/315 of the Generalitat de Catalunya.
S.C. and A.P. acknowledge the financial support of INAF through the 
PRIN MIUR 2007: \lq{Multiple stellar populations in globular clusters}, 
and ASI grant ASI-INAF I/016/07/0, 
the financial support from the Italian Theoretical Virtual
Observatory  Project as well as the help provided by P. Manzato, M.
Molinari, and F. Pasian in improving and maintaining the BaSTI database.
This research has made use of NASA's Astrophysics Data System Abstract
Service and the SIMBAD database operated at CDS, Strasbourg, France.

\end{document}